\documentclass[12pt]{article}
\usepackage{setspace}
\usepackage{latexsym}
\usepackage[dvips]{graphicx,color}
\usepackage{graphics}
\usepackage{graphicx}
\graphicspath{{Graphics/}} 
\usepackage[space]{grffile}
\usepackage{pstricks}
\usepackage{epstopdf}
\usepackage[T1]{fontenc}
\usepackage{amsmath, amssymb,slashed,url}
\usepackage{amsfonts}
\usepackage{hyperref}
\usepackage[utf8]{inputenc}

\newcommand{\dsl} {\kern.06em \hbox{ \raise.15ex \hbox{$/$} \kern-.56em\hbox{$\partial$}}}

\newcommand{\ZZ}{{\rm \kern 0.275em Z \kern -0.92em Z}\;}

\newcommand{\beq}{\begin{equation}}
\newcommand{\eeq}[1]{\label{#1}\end{equation}}
\newcommand{\bea}{\begin{eqnarray}}
\newcommand{\eea}[1]{\label{#1}\end{eqnarray}}
\renewcommand{\Re}{\mbox{Re}\,}

\def\Tr{\mbox{Tr}\,}

\newcommand{\bc}{\begin{center}}
\newcommand{\ec}{\end{center}}
\def\Tr{\mathop{\rm Tr}\nolimits}

\title{Cylinder Transition Amplitudes in Pure AdS3 Gravity}
\author{Alan Garbarz$^{a}$~\footnote{E-mail: \href{mailto:alan@df.uba.ar}{alan@df.uba.ar}}, Jayme Kim$^{b}$~\footnote{E-mail: \href{mailto:jk2943@nyu.edu}{jk2943@nyu.edu}}, Massimo Porrati$^{c}$~\footnote{E-mail: \href{mailto:massimo.porrati@nyu.edu}{massimo.porrati@nyu.edu}}}
\date{%
    $^a$ Physics Department \& IFIBA-Conicet, University of Buenos Aires\\ 
Pabell\'on 1, Ciudad Universitaria, 1428 Buenos Aires, Argentina\\[2ex]%
    $^b$ Department of Physics and Astronomy, Montclair State University \\ 1 Normal Ave, Montclair NJ 07043, USA\\[2ex]%
    $^c$ Center for Cosmology and Particle Physics, \\ Department of Physics, New York University, \\726 Broadway, New York, NY 10003, USA\\[2ex]%
}

\begin{document}
\maketitle



\begin{abstract}
A spacelike surface with cylinder topology can be described by various sets of canonical variables within pure AdS3 gravity. Each is made of one real coordinate and one real momentum. The Hamiltonian can be either 
$H=0$ or it can be nonzero and we display the canonical transformations that map one into the other, in two relevant  cases.  In a choice of canonical coordinates, one of them is the cylinder aspect $q$, which evolves nontrivially in time. The 
time dependence of the aspect is an analytic function of time $t$ and an ``angular momentum'' $J$. By analytic continuation 
in both $t$ and $J$ we obtain a Euclidean evolution that can  be described geometrically as the motion of a cylinder inside 
the region of the 3D hyperbolic space bounded by two ``domes'' (i.e. half spheres), which is topologically a solid torus. We 
find that for a given $J$ the Euclidean 
evolution cannot connect an initial aspect to an arbitrary final aspect; moreover, there are 
infinitely many Euclidean trajectories that connect any two allowed initial and final aspects. We compute the transition 
amplitude in two independent ways; first by solving exactly the time-dependent Schr\"odinger equation, then by summing in a sensible way all the saddle contributions, and we discuss why both approaches are mutually consistent.
\end{abstract}

\newpage
\section{Introduction}

The study of quantum transitions between geometries in General Relativity is a central piece in the construction of a complete Quantum Gravity.  It is however a prohibitively difficult matter in four dimensions but theories of gravity in lower dimensions may be less intractable.  A particularly simple yet rich playground consist of GR with a negative cosmological constant in three dimensions. This {\em AdS$_3$ pure gravity} has no local degrees of freedom, which simplifies greatly the constraints and dynamical equations.  In fact, explicit solutions of the general classical dynamics of point particles 
exist for zero, positive and negative cosmological constant~\cite{deser}. What makes AdS$_3$ special is that it also possesses ``standard''  black hole solutions \cite{Banados:1992wn, Banados:1992gq} as well as topologically non-trivial black holes \cite{Brill:1995jv, Aminneborg:1997pz, Brill:1998pr}.

Research on the global dynamics and quantization of AdS$_3$ gravity was initiated by many authors in the 1980's  
\cite{Martinec:1984fs, Achucarro:1987vz, Witten:1988hc, Witten:1989sx, Martin:1989mb, Carlip:1989nz}. In particular, a 
remarkable analysis of dynamics was developed for closed initial surfaces by Montcrief \cite{Moncrief:1989dx}. 
As a concrete example, ref~\cite{Moncrief:1989dx} studied in details  the dynamics of a toroidal initial surface. Roughly, the point made there was that the phase space of the theory is given by the cotangent space of the Teichm\"uller space of the initial spacelike surface so that a sensible quantization could be stated in terms of wave functions on Teichm\"uller space by geometric quantization (see also \cite{Witten:1988hc}).  This work was soon followed by other ones, that also 
considered the case of closed spacelike surfaces and in particular the torus \cite{Hosoya:1989yj,Carlip}.

In this work we are interested in the case where the initial/final surfaces are open and have black hole asymptotics. 
A framework for the canonical quantization of pure gravity in such geometries was developed in~\cite{kp15}. While that
reference focused on the intrinsic dynamics of the moduli characterizing a genus-$g$ initial surface, that can be said to 
describe the interior of a black hole, the present paper will study the ``redundant'' dynamics of the exterior.

Specifically, here we shall consider spacelike surfaces with cylindrical topology,  where the ends are at infinity, as in the 
maximally extended BTZ black hole \cite{Banados:1992gq}. This can be seen as a half annulus in 
the complex upper half-plane described by the complex coordinate $z$, 
by making the identification $z\sim \lambda z$, with 
$\lambda>1$. A fundamental region is given by $1\leq |z| <\lambda$ and $0<\text{arg}(z)<\pi$. As it will turn out,
$\theta=\log\lambda$ plays a central role in the present work, because it is a coordinate for the Teichm\"uller space of the cylindrical equal-time surfaces that we will be considering. 

We start our paper by describing the well-known Hamiltonian dynamics of AdS$_3$ gravity, for a generic initial surface $\Sigma$, following \cite{Scarinci:2011np}. In a particular gauge, the natural dynamical variables are the conformal (or complex) structure and the extrinsic curvature of $\Sigma$. The dynamical equations can be solved and the evolution of the classical variables gives a hyperbolic three-manifold. When we consider the case where $\Sigma$ has a cylindrical topology, the complex structure is related to $\theta$ and the classical evolution of this coordinate can be read from the induced metric at some other time. In this manner, we obtain the classical evolution of the complex structure.   

Next we consider different sets of canonical variables describing the dynamics of the cylinder. They are generically given by two real numbers, one of which 
is related to the cylinder aspect while the other can be thought of as an angular momentum.  We also discuss the generating functions responsible for the canonical transformations between sets of canonical coordinates. These functions can be related to boundary terms at $\Sigma$ in the action, but we will not analyze such boundary terms in detail. Instead we specialize our discussion to a particular set of coordinates, where the generalized coordinate $q$ is the aspect of the cylinder while the other is an ``angular momentum'' $J$, related to the extrinsic curvature. Both undergo an
evolution determined by a specific Hamiltonian, which we also give.  We also compute the exact quantum 
evolution between different surfaces exactly solving the Schr\"odinger equation for the given Hamiltonian.

The transition amplitude can also be computed in the path integral approach by considering the Euclidean evolution 
interpolating between two given cylinder aspects (i.e.\ two different $q$'s). Such evolution can be visualized as a cylindrical
surface moving inside a hyperbolic three-manifold. The cylinder then evolves without topology change, but changing its aspect. One surprising finding is that there is an upper bound on the ratio of the two $q$'s, meaning that, given an initial aspect, there is only a bounded set of reachable aspects. Also, given two aspects that can be joined by an interpolating hyperbolic three-manifold there are infinitely many such Euclidean saddles (which are 
roughly parametrized by the angular momentum $J$).       

Taking into account these observations, we compute the path integral in the saddle-point approximation by summing over 
the infinite saddles. We conclude by comparing the result of the Euclidean path integral computation with the exact solution
of the Schr\"odinger equation.

\section{Classical dynamics of the cylinder}

We begin by partially fixing the gauge in a few steps. First, let us write a time foliation of spacetime as $M=\mathbb{R}\times \Sigma$ and choose a time gauge where the lapse function is 1 and the shift is zero ($N_t=1$ and $N_{\text{shift}}=0$). This is different from the choice of  \cite{Moncrief:1989dx} where each equal-time surface has constant mean curvature so that time equals the mean curvature. We cannot make this choice because constant mean-curvature foliations only work in general when the spacelike surfaces are closed \cite{Krasnov:2005dm}.  Next we choose isothermal coordinates on each chart on $\Sigma$. This means that locally 
\begin{equation}\label{sigmametric}
ds^2 = e^{2\phi} |dz|^2,\quad \text{on}\; \Sigma  ,
\end{equation}
with $\phi$ an arbitrary real function of $z$ and $\bar{z}$. There are infinite choices of isothermal coordinates related to
each other by holomorphic coordinate transformations. We will come back to this important point soon.

In the Hamiltonian formulation of GR one needs to give also the extrinsic curvature (or second fundamental form on $\Sigma$)
\begin{equation}
II=\frac{1}{2} \left( qdz^2 + \bar{q}d\bar{z}^2\right)+H e^{2\phi}|dz|^2 .
\end{equation}
Here $q$ is a quadratic differential (not to be confused with the canonical coordinate!) and $H$ is the mean curvature. The Codazzi and Gauss constraints read respectively,
\begin{equation}
\bar{\partial}q=e^{2\phi} \bar{\partial}H,\quad 4\partial\bar{\partial}\phi=e^{2\phi}(H^2-\Lambda)-e^{-2\phi} |q|^2 ,
\end{equation}
with $\Lambda=-1/\ell^2$ which we set to $-1$ from now on. We can actually choose $H=0$ for $\Sigma$, implying that $\bar{\partial}q=0$, so $q$ is a holomorphic quadratic differential. The Gauss constraint is a Liouville-like equation for $\phi$
\begin{equation}
4\partial\bar{\partial}\phi=e^{2\phi}-|q|^2e^{-2\phi} ,
\end{equation}
which has a unique solution for a given $q$ (see \cite{Krasnov:2005dm}). 

The remaining GR equations give the evolution of the initial metric (\ref{sigmametric}) and of the extrinsic curvature. They can be solved in our gauge to give
\begin{equation}\label{metric}
ds^2=-dt^2+\cos^2{t}\,e^{2\phi} |dz+\tan{t} \,\bar{q}\, e^{-2\phi} d\bar{z}|^2  .
\end{equation} 
In short, the choice of isothermal complex coordinate  $z$ and  of the holomorphic quadratic differential $q$ uniquely determine the evolution and thus the phase space. However this is not strictly correct since, as we anticipated, the complex coordinates $z$ and $w=f(z)$ (with $f$ holomorphic) are both equally admissible. In other words, metrics that
can be transformed into each other by a diffeomorphism on $\Sigma$ that maintains the isothermal gauge should not be
considered different; this implies that we must quotient by these holomorphic diffeomorphisms. What is left is the space of complex structures on $\Sigma$, namely Teichm\"uller space \footnote{This is also the space of conformal structures. Technically we are quotienting by diffeomorphisms homotopic to the identity. If we quotient by all the (orientation-preserving) diffeomorphisms, then we do not have the Teichm\"uller space but the Riemann moduli space. The latter is the former divided by the mapping class group.}. The phase space is then given by the complex structure and the extrinsic curvature (see \cite{Scarinci:2011np} for further discussions on the phase space structure). In what follows we will take $\Sigma$ to be an open, genus zero surface with two holes. 

\subsection{$\Sigma$ as annulus and as cylinder}\label{AnnulusCylinder}

Considering that $z$ lives on the upper half plane, then we should take 
$$z \sim \lambda_0\, z,$$ 
with $\lambda_0>1$. This gives a fundamental region in the upper-half plane, bounded by two semicircles of radius 1 and $\lambda_0$, which are identified. The asymptotic boundaries are at arg$(z)=0$ and arg$(z)=\pi$. The identification can be written as an SL$(2,\mathbb{R})$ modular transformation given by diag$(\sqrt{\lambda_0},1/\sqrt{\lambda_0})$, whose
 trace is a good parameter for labeling a point in Teichm\"uller space, which is tantamount to say that $\lambda_0$ is a good modulus.  

We can equivalently consider a holomorphic transformation $w=\log z$, which maps the annulus to a rectangle of fixed 
height $\pi$ and width $\theta_0:=\log\lambda_0$ in the complex plane. The identification in the annulus maps to identifying
 Re$(w)=0$ with Re$(w)=\theta_0$, while the asymptotic boundaries are at Im$(w)=0$ and Im$(w)=\pi$.  The aspect of the rectangle is then related to $\theta_0$, that we can therefore use as a coordinate of Teichm\"uller space. 

As a technical but relevant note, we should mention that when the surface has boundaries, then there are two possible 
definitions of Teichm\"uller space, depending on the behavior at the boundaries of the diffeomorphisms that 
one uses to take 
quotient. The point we want to stress for the cylinder (or equivalently for the annulus) 
is that there are two standard definitions
of the Teichm\"uller space. In one, 
Teichm\"uller space is infinite dimensional while in the other the \textit{reduced Teichm\"uller space} is one-dimensional\footnote{We thank Scott Wolpert for clarifications on this issue and for suggesting the book by Nag \cite{Nag}}. We are considering this latter case, which for the cylinder topology is in fact equal to the Riemann moduli space \cite{Nag}.   

\subsection{Solving the constraints}

We already mentioned that the Codazzi equation implies that $q$ is actually holomorphic. Then, with a slight abuse of
 notations we need $q=q(z) dz^2$ to be invariant under the identification $z\sim \lambda_0 z$.  In the annulus picture, 
\begin{equation}\label{q}
q|_{\text{annulus}}=\frac{\bar{J}}{a^2}  \frac{dz^2}{z^2},\quad J\in\mathbb{C}, \quad a:=\frac{\pi}{2K(|J|^2)}
\end{equation}
is evidently invariant under $z\mapsto \lambda_0 z$. $a$ is a real number that is there just to make future equations look simpler. The important point is that $q$ depends on a complex parameter $J$ \footnote{The parameter $J$ controls the extrinsic curvature. To make contact with a BTZ black hole, it should be equal to $ i r_-/r_+$.}(we used $\bar{J}$ here because in 
eq.~(\ref{metric}) it is $\bar{q}$, hence $J$  that appears). 

In the cylinder coordinate $w=\log z$, $q$ must be translation invariant. It is given by
\begin{equation}
q|_{\text{cylinder}}=\frac{\bar{J}}{a^2}  {dw^2} .
\end{equation}
This holomorphic quadratic differential gives the Gauss constraint
\begin{equation}
4\partial\bar{\partial}\psi=e^{2\psi}-\frac{|J|^2}{a^4}e^{-2\psi} , \qquad \psi :=\phi +\Re w .
\end{equation}
Let us write 
\begin{equation}
w=x+i y ,
\end{equation}
so the solution to the Gauss equation is
\begin{equation}
\psi(w,\bar{w})=\psi(y)=-\frac{1}{2}\log \left(a^2 \text{sn}^2(y/a) \right) ,
\end{equation} 
where the elliptic sine has parameter $|J|^2$. 

\subsection{Exact classical evolution}

Now that we have solved the constraints, it is time to understand what is the evolution in the constrained surface, i.e. in the physical phase space. Let us define the Beltrami differential $\mu$
\beq
\mu:=\tan t \, I e^{-2\psi},\quad I:=\frac{J}{a^2} ,
\eeq{2.3.1}
or in other words $\mu=\tan t\,\bar{q} \,e^{-2\psi}$. This means that (\ref{metric}) can be written as
\beq
ds^2=-dt^2+\cos^2{t}\,e^{2\psi} |dw+\mu \,d\bar{w}|^2 .
\eeq{2.3.2}
It is clear that at $t=0$ the complex coordinate is $w$ and the complex structure is $\theta_0$. The question is what is the 
complex structure at an arbitrary time $t$. Notice that the metric~(\ref{2.3.2}) is not global because it terminates  in a 
coordinate singularity on the surface  $t=T(w,\bar{w})\leq \pi/2$, which is where
the determinant of the spacelike metric vanishes
\beq
h:=\det h_{ab}= \cos^4 t \, e^{4\psi} \left(1-|I|^2 \tan^2 t \, e^{-4\psi} \right)^2=0 , \qquad {a,b=1,2}.
\eeq{sing}
Notice that the trace of the extrinsic curvature  diverges at the coordinate singularity
\beq
K= K_{ab}h^{ab} = {1\over \sqrt{h} } {\partial \over \partial t} \sqrt{h} = 
-2\tan t {1+|I|^2 e^{-4\psi} \over 1-|I|^2 \tan^2 t \, e^{-4\psi}} .
\eeq{sing-k}
Since $K$ plays the role of time in the Moncrief parametrization~\cite{Moncrief:1989dx},  eq.~(\ref{sing-k}) shows that 
his time coordinate cannot cover a spacetime region extending beyond the singularity.

At time $t$ the induced metric on the spacelike surface is conformal to  $|dw+\mu \,d\bar{w}|^2$, so we are looking for a complex coordinate $X$ such that $|dw+\mu \,d\bar{w}|^2 \propto |dX|^2$ at time $t$. 
The equation to solve is the Beltrami differential equation 
\begin{equation}\label{Beltramidiffeq}
\bar{\partial}X=\mu\partial X .
\end{equation}
The unique normalized\footnote{Normalized here means that it fixes the points $0$, $1$, and $\infty$. Also, $\sigma_y$ can be written in terms of a complex incomplete elliptic integral of the third kind.} solution on the upper-half plane is
\begin{equation}
X=e^{A (x+i \sigma_y )}, \quad \sigma_y:=\int_0^y dy'\frac{1-\mu}{1+\mu} ,
\label{quasimap}
\end{equation}
with,
\begin{equation}\label{A}
A:=\frac{\pi}{\text{Re}\,\sigma_\pi} .
\end{equation}
Of course, for the cylinder we have 
\begin{equation}
w_t=A( x + i \sigma_y) .
\end{equation}
The real number $A$ has the virtue of making the map $w\mapsto w_t$ bijective on the cylinder of height $\pi$ when $\sigma_y$ is real (namely when $J\in\mathbb{R}$), implying that the new aspect only  depends on the width given by $A \theta_0$. In other words the new point in 
Teichm\"uller space is,
\begin{equation}\label{theta}
\theta(t):=A\,\theta_0
\end{equation}
Note that $A$ depends both on $t$ and on $J$. 

If  Im$(J)\neq 0$, then the rectangle $[0,\theta_0]\times [0,\pi]$ where $w$ lives is mapped to a generalized quadrilateral in 
the complex plane, which is not another rectangle. One way to understand the new conformal structure associated to this 
generalized quadrilateral is to first conformally map it onto the upper half plane, then apply a conformal Christoffel-Schwarz
 map to a new cylinder and finally read the aspect \cite{Gardiner}. However this procedure requires to know the first 
 conformal map onto the upper-half plane (which exists by the Riemann Mapping Theorem). A much more direct way to 
 read the conformal structure is to realize that the initial Fuchsian group implementing the quotient on $\mathbb{H}^2$ is 
 given by $z\sim \lambda_0 z$ or, equivalently, the PSL$(2,\mathbb{R})$ matrix $M=$diag$(\sqrt{\lambda_0},
 1/\sqrt{\lambda_0})$. The new complex structure comes from a quasi-conformal deformation of such matrix. The new 
 Fuchsian group acts as $M'\cdot X(w,\bar{w}) =X \circ M\cdot \exp (w) $, meaning that $X\sim e^{A \log\lambda_0} X$, and
  thus equation (\ref{theta}) still holds.

\subsection{Canonical coordinates and transformations}

So far we have focused on the evolution of the cylinder aspect, thought (somewhat implicitly) as the generalized coordinate $q$. We would like to show that, as in the case of the torus studied in \cite{Carlip}, there is freedom in choosing a pair of canonical coordinates, with each pair governed by a well defined Hamiltonian.
  
We start from the canonical pair $(p,q)$ defined as \footnote{We hope there is no confusion between this canonical coordinate $q$ and the holomorphic quadratic differential $q=q(z)dz^2$.}
\beq
p:=2\pi \Re I, \qquad q:=\theta_0=\log \lambda_0 .
\eeq{ct1}
They do not evolve in time, so the corresponding Hamiltonian is $H=0$.
We want to find the canonical transformation that maps $q$ and $p$ to the new coordinate $Q$ that evolves as in (\ref{theta}).
\beq
Q={\pi \over \Re \sigma_\pi} q, \qquad \sigma_y=\int_0^y ds {1-\tan t I e^{-2\psi(s)} \over 1+ \tan t I e^{-2\psi(s)}},
\qquad 0 \leq y \leq \pi .
\eeq{ct2}
The function $\psi(y)$ obeys the equation $d^2\psi /d^2y= \exp(2\psi) -I\bar{I} \exp(-2\psi)$, so $\Re \sigma_\pi$ is a 
function of $p=2\pi \Re I$, $t$ and $I\bar{I}$, \underline{but not of q}. So we can write the equation relating old and new
coordinates as
\beq
q= {Q \over \pi} \Re \sigma_\pi(p,t,I\bar{I}).
\eeq{ct3}
Now it becomes very easy to find the canonical transformation, if we use the correct set of coordinates.
By definition a canonical transformation leaves the canonical one-form invariant up to an exact differential
\bea
pdq -Hdt + dF &=& d(pq)- qdp -Hdt +dF = PdQ -H'dt   \qquad\Longrightarrow  \nonumber  \\
q={\partial \Phi \over \partial p}, \qquad P&=& {\partial \Phi \over \partial Q},
\qquad H'=H- {\partial \Phi \over \partial t}, \qquad \Phi\equiv pq +F .
\eea{ct4}
The independent coordinates in the function $\Phi$ are $p$, $Q$ and $t$. Because eq.~\ref{ct3} expresses $q$ in
terms of the independent coordinates it can be integrated in $p$ to give
\beq
\Phi(p,Q,t)={Q \over \pi} \int_0^p dp'   \Re \sigma_\pi(p',t,I\bar{I}) + f(Q,t,I\bar{I}).
\eeq{ct5}
We do not have a good argument to fix $f(Q,\tau,I\bar{I})$, the $p$-independent part of $\Phi$, neither we have an argument that
says it cannot be zero, so we will choose $f=0$\footnote{Note that different choices of $f$ amount to $Q$-dependent translations in $P$, which are canonical transformations.}. This gives the new momentum $P$ and the new Hamiltonian $H'$ as
\beq
P={1 \over \pi} \int_0^p dp'   \Re \sigma_\pi(p',t,I\bar{I}), \qquad H'=-{Q \over \pi} \int_0^p dp'   
{\partial \Re \sigma_\pi\over \partial t}(p',t,I\bar{I})  .
\eeq{ct6}

\subsection{Exact quantum evolution}

The classical Hamiltonian in eq.~(\ref{ct6}) is of the general form $H= v^a(p,t)q_a$. Canonical quantization is done by 
the replacement $q_a\rightarrow i\partial/\partial p^a$. By demanding that  the Hamiltonian be Hermitian we get the
exact time-dependent Schr\"odinger equation in the momentum basis
\beq
{\partial\over \partial t} \psi(p)= \left[ v^a(p,t) {\partial \over \partial p^a} + {1\over 2} \partial_a v^a(p,t)  \right]\psi(p).
\eeq{mm4}
In the absence of the $\partial_av^a$ term, the exact general solution is standard, see e.g. Chapter 2 of ~\cite{arnold}. 
In the presence of that term we proceed as follows.
First we define $P^a(p,t)$, the integral of the characteristic equation 
\beq
{\partial P^a \over \partial t}= -v^a(P,t) 
\eeq{mm5}
obeying the initial condition 
$P^a(p,0)=p^a$.
This integral is then used to define $\hat{P}^a(p,t)$ as the ``level set'' 
\beq
P^a(\hat{P}(p,t),t)=p^a.
\eeq{mm6}
By differentiating eq.~(\ref{mm6}) w.r.t. $t$ we get
\beq
{\partial P^a \over \partial \hat{P}^b} {\partial \hat{P}^b \over \partial t} - v^a(p,t)=0 \Rightarrow
{\partial \hat{P}^a \over \partial t}= {\partial \hat{P}^a \over \partial p^b} v^b(p,t).
\eeq{mm7}
Next we define $H_a^{\; b}=\partial_a \hat{P}^b$ and its inverse $K_a^{\; b} $ ($K_a^{\; b} H_b^{\; c}=\delta_a^{\; c}$). 
The time derivative of $H\equiv \det H_a^{\; b}$ is
\bea
{\partial \over \partial t} H &=&  H K_a^{\; b} {\partial \over \partial t} H_b^{\; a}  = H K_a^{\; b} \partial_b \left[ v^c(p,t)\partial_c \hat{P}^a(p,t) \right] = H K_a^{\; b} \partial_b v^c(p,t) H_c^{\; a} + H K_a^{\; b}v^c(p,t)\partial_c H_b^{\; a} \nonumber \\
&=& 
H \partial_c v^c(p,t) + v^c(p,t)\partial_c H.
\eea{mm8}
Equations~(\ref{mm7},\ref{mm8}) allow us to write the general solution to eq.~(\ref{mm4}) obeying the initial condition
$\psi(p,0)=\phi(p)$  as
\beq
\psi(p,t)= H^{1/2} \phi(\hat{P}(p,t)).
\eeq{mm9}
It is easy to check that the time evolution preserves the norm
\beq
\int [dp] \psi^*(p,t)\psi(p,t)=\int [dp] \det \partial_a \hat{P}^b \phi^*(\hat{P}(p,t))\phi(\hat{P}(p,t))=\int d[\hat{P}]\phi^*(\hat{P})\phi(\hat{P}).
\eeq{mm10}
In fact, we could have gotten the factor $H^{1/2}$ in~(\ref{mm9}) precisely by requiring conservation of the norm.

\subsection{Complex structure of a BTZ initial surface}

We will compute later a transition amplitude between cylinders with different aspect. We will do this by means of a Euclidean semiclassical path-integral approximation. We first note that the function $A$ in (\ref{A}) is analytic in $t$ and $J$ so that we can Wick rotate these parameters. We will come back to this point in the next Section while for now we just
want to understand the complex structure of a Euclidean BTZ, in order to make contact with this standard
 black hole geometry. 

The Euclidean BTZ at fixed time is \footnote{The usual BTZ coordinates $(r,\tau,\varphi )$ only cover one exterior region $r>r_+$, however one needs two identical patches to cover the whole annuli. We will take this into account later.} 
\begin{equation}
ds^2|_{\tau=0}=\frac{r^2}{(r^2-r_+^2)(r^2+r_-^2)}dr^2+r^2 d\varphi^2  .
\end{equation}
This 2D metric can be put in isothermal coordinates as follows. First take\footnote{Here $F$ is the incomplete elliptic integral of first kind and $K$ is the complete elliptic integral of first kind.}
\begin{equation}
\alpha:=a_\kappa F(\arcsin(r_+/r),-\kappa^2), \qquad \text{with}\quad \kappa:=\frac{r_-}{r_+},\quad 
a_\kappa:=\frac{\pi}{2 K(-\kappa^2)} ,
\end{equation}
where $\alpha\in (0,\pi/2]$. Then define the complex coordinate 
\begin{equation}
z:= e^{\log\lambda_0 \frac{\varphi}{2\pi} +i \alpha},\qquad \lambda_0=e^{2\pi a_\kappa r_+} .
\end{equation}
Note that the parameters $r_+$ and $\kappa$ define the Teichm\"uller parameter $\lambda_0\in (1,\infty)$ of the annulus $\mathbb{H}/\sim$, with $z\sim\lambda_0 z$. Also note that this complex coordinate $z$ covers only one portion of the annulus (up to $\alpha=\pi/2$) but it is immediate to extend it to $\alpha=\pi$. 

\begin{figure}[h!]
\centering
    \includegraphics[width=.5\linewidth]{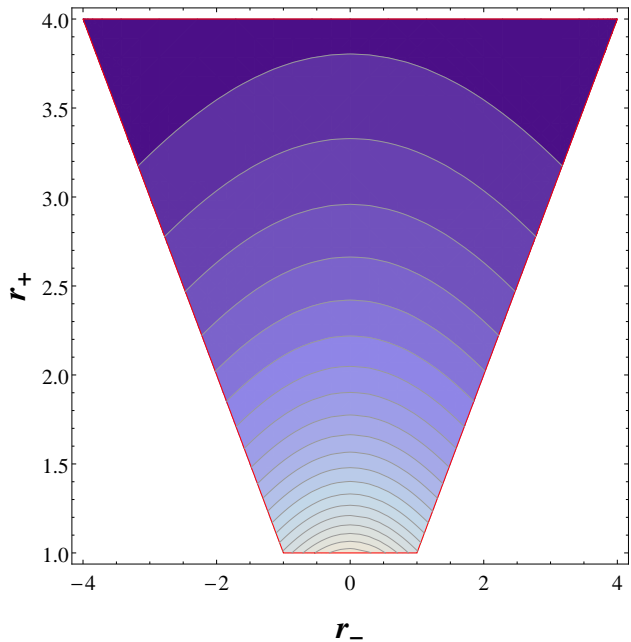} \\
\caption{Numerical plot of level curves of $\lambda_0$ as functions of $r_+$ and $r_-$.}
\label{contourBTZ}
\end{figure}

The different values of $\lambda_0$ in terms of $r_+$ and $r_-$ can be visualized by a contour plot such as the one 
given in Figure \ref{contourBTZ}.
We should stress that a given $\lambda_0$ does not identify a BTZ black hole but a family of such black holes. The remaining data needed is of course the extrinsic curvature of the initial surface.

\subsection{Analysis of the Euclidean evolution} 

Once Wick rotated, the function $A$ depends on $\tanh\tau$ and on the Euclidean continuation of $I=J/a^2$,  which comes from the Euclidean continuation of $J$. This is actually implemented as $J\rightarrow i J, \bar{J}\rightarrow i \bar{J}$, in order to change the $-q \bar{q} e^{-2\psi}$ term in the Gauss constraint to $+q \bar{q} e^{-2\psi}$, which is the correct one for the Euclidean setting. We henceforth still call them $J$ and $I$. Now $I=J/a^2$ but with $a={\pi \over 2 K(-|J|^2)}$ (compare with eq. (\ref{q})). Accordingly, elliptic functions now have parameter $-|J|^2$. 

\begin{figure}[h!] 
\centering
    \includegraphics[width=.8\linewidth]{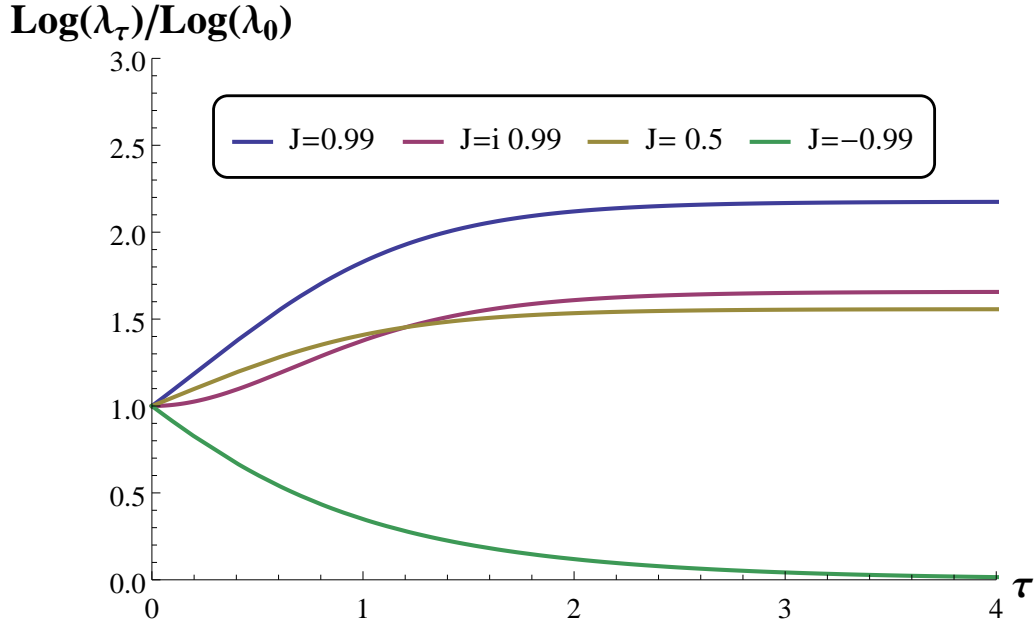} \\
 \caption{The curves of fixed $J$ in (\ref{A}) as a function of $\tau$. We see from the intersection at some $\tau$ that there is more than one saddle  (more than one $J$) for each final structure $\log\lambda_\tau$. }
 \label{fig:lambda}
\end{figure}

Since $\mu$ needs to satisfy\footnote{This is a result  of Teichm\"uller theory but it is  motivated 
more directly by noticing that the metric is invertible only 
when this condition holds.} $||\mu||_\infty < 1$, then the inequality  $|J \tanh\tau|< 1$ holds. This in turns implies that there is a bound on how much the $\theta_t=\log\lambda_\tau$ parameter can change compared to $\theta_0=\log\lambda_0$. The range of $\theta_\tau$ is given by the minimum and maximum values of $A$ for fixed $\tau$ in (\ref{A}), see Figure \ref{fig:lambda}. Actually, the minimum value is zero, because the function $\sigma_\pi$ diverges as it approaches $J=-\coth\tau$. 

It is an amusing fact that there is a maximum value of $\theta$ given $\theta_0$, which corresponds to $J=\coth\tau$.  For 
example, for infinite Euclidean time the maximum value is approximately $2.18844 \theta_0$. For finite time, the maximum 
may be larger than this number, but still finite. At any time, this bound on the ratio $\theta/\theta_0 $ of Teicm\"uller parameters is \textit{more stringent} than the purely geometric bound coming from studying K-quasiconformal maps (see the Conclusions).   So we seem to have found a new dynamical selection rule.

\begin{figure}[h!]
\centering
    \includegraphics[width=\linewidth]{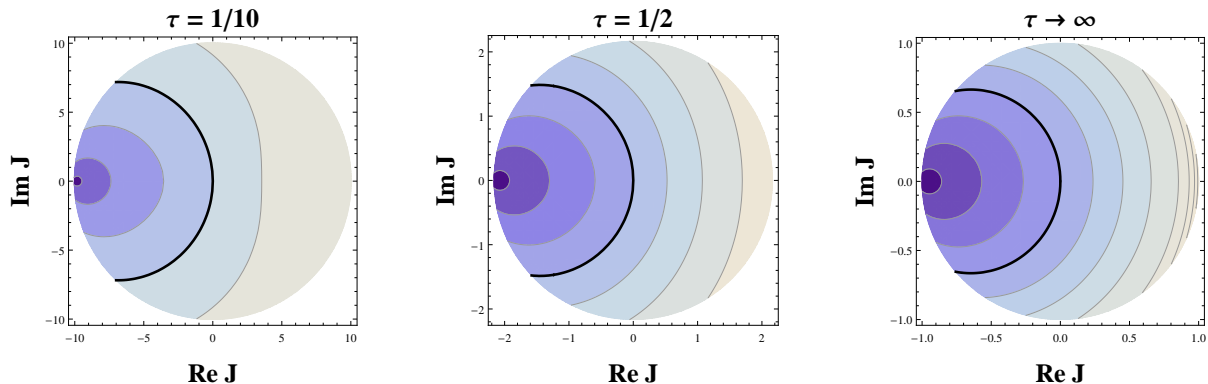} \\
\caption{Three contour plots of (\ref{A}) for times $\tau=0.1, .5, \infty$. The black thick curve in each plot is the level curve of value $1$, formed by instantons which do not change the complex structure. The disks have radius $\coth\tau$. It is evident that every level curve intersects the real line only once.}
\label{contourgrid}
\end{figure}

Note that even when $\lambda_\tau$ is fixed there are infinite values of $J$ that give the same value of $A$. To see the 
root of this issue we should remember that $\mu$ depends on $J$, but from Teichm\"uller theory we know that different 
Beltrami coefficients $\mu$, that is different $J$'s may give the same complex structure. In other words, it is the class 
$[\mu]$, rather than the parameter $\mu$, which defines the complex structure at time $\tau$.  This is related to the discussion at the end of eq.~(\ref{AnnulusCylinder}), since the choice of reduced Teichm\"uller space implies that the quadratic differential $q$ should be real on the boundary in order to identify it with the fibre of the cotangent phase space, so that  $J$ should also be real. This is from a 2-dimensional perspective, but from the 3-dimensional interpolating manifold, there is no reason to require a real $J$, yet it is only $\Re{J}$ that parametrizes unambiguously the final complex structure.  This point can be made more clear: the level sets of $A$ for given $\tau$ can be visualized in Figure \ref{contourgrid}. Each level curve of $A$ intersects the real $J$ line only once.    


\newpage

\section{Semiclassical transitions between conformal structures}

\begin{figure}[h!]
\centering
{\includegraphics[width = .7\linewidth]{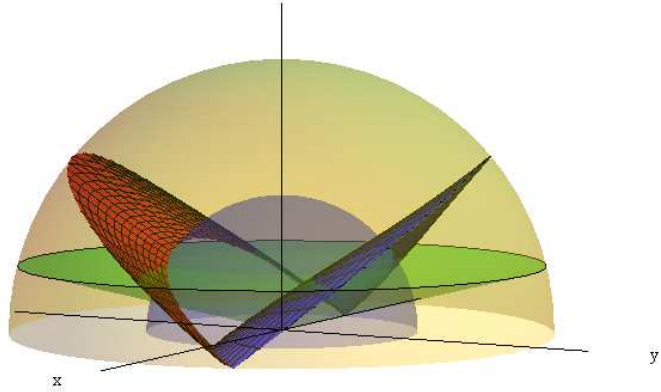}}
\caption{Fundamental region between two domes and two constant-$\tau$ surfaces, together with the regularizing cone (in green). The parameter $J$ is set to $i/2$. The integration is to be taken between the domes as well as between the two equal-time surfaces and above the cone. This image was obtained from the coordinate transformation described in the Appendix.}
\label{fundamentalregion}
\end{figure}

In this Section we begin the computation of a Euclidean quantum transition from an aspect $\theta_0$ to some other aspect 
$\theta$ at time $\tau$. We want to compute $\langle \theta,\,\tau|\theta_0,\,0\rangle $, and we do it by the path-integral 
method and considering only the semiclassical contribution given by the on-shell action.  The exact quantum evolution was computed at the end of the previous section in the ``constrain first then quantize'' 
approach. The path integral quantization instead can compute a wave function defined on unconstrained 3-metrics,
 that formally satisfies the Wheeler-DeWitt equation~\cite{haha},
but here we will work to the lowest order in the semiclassical approximation so the
``constrain first'' and the path integral approaches should agree to lowest order. It is 
thus meaningful and maybe instructive  to compare the results of the two approaches.
We need first of all to compute the regularized on-shell action of the hyperbolic Euclidean interpolating geometries discussed in the previous Section. 
The metric is now 
\begin{equation}\label{Euclideanmetric}
ds^2=d\tau^2+\cosh^2{\tau}\,e^{2\psi} |dw+\tanh{\tau} \,I\, e^{-2\psi} d\bar{w}|^2  ,
\end{equation} 
while the manifold is $\mathbb{H}^3/\Gamma$, where $\Gamma$ is generated by $\vec{X}\sim \Lambda \vec{X}$, with $\Lambda$ a complex number depending both on  $\lambda_0$ and $J$. For example, for the Euclidean BTZ, $J=i\kappa$ and  $\Lambda=e^{2\pi r_+(1+J)}$. A natural  fundamental region for $\mathbb{H}^3/\Gamma$ is the region between two domes, one of radius one and the other of radius $|\Lambda|$. The reader should have Figure \ref{fundamentalregion} in mind, where the blue and red surfaces represent the initial and final cylinders, while the green surface is a regulating surface, which will be defined in the following pages

\subsection{Euclidean saddle}
\subsubsection*{The case with $\mu=0$}
Let us learn how to regularize the Euclidean action with boundaries at finite time by looking first 
at the case $\mu=0$. Let us consider the metric
$ds^2=d\tau^2+\frac{\cosh^2\tau}{\sin^2 y} (dy^2+d\varphi^2)$ 
where $y\in(0,\pi/2)$ and $\varphi\sim\varphi+2\pi$. We want to integrate from $T_1$ to $T_2$, and consider the possible regularization $\sin y \rightarrow \sin\epsilon \cosh\tau$. In order to understand this, let us go to canonical coordinates in hyperbolic space:
\begin{eqnarray}
X&=&e^{\theta_0 {\varphi\over 2\pi}} \cos y , \nonumber\\
Y&=&e^{\theta_0 {\varphi\over 2\pi}} \sin y ,\tanh\tau \nonumber\\
Z&=&e^{\theta_0 {\varphi\over 2\pi}} { \sin y \over  \cosh \tau} .
\end{eqnarray}
Now we see that the initial/final surfaces are planes such that $\sinh\tau=Y/Z=\text{constant}$, while a constant $y$ represents cones with axis $X$: 
$$\left\{\cot y=X/\sqrt{Y^2+Z^2}=\text{constant}\right\}.$$
The regularization means taking a wide cone along the $Z$ axis such that $Z=\sin\epsilon\sqrt{X^2+Y^2+Z^2}$, and at each $\tau$ integrate $y$ inside this cone starting from $y\simeq \sin y=\sin\epsilon \cosh\tau$.
We can see that the large $|T|$ limit is bounded from above by the constraint $\cosh\tau\leq (\sin\epsilon)^{-1}$. We will soon show that this allows us to recover Krasnov's result in \cite{Krasnov} for the entire volume between the domes.

The different contributions to the Euclidean action are: the volume,
\begin{eqnarray}
V:=\frac{\theta_0}{\tan^2\epsilon} \left(\tan\epsilon \sinh\tau\sqrt{1-\tan^2\epsilon \sinh^2\tau}+ \arcsin\left(\tan\epsilon\sinh\tau\right)\right)\large |_{T_1}^{T_2}  
\end{eqnarray}
the Gibbons-Hawking term and area term for the regulating cone\footnote{The conventions are: $$K_{ab}=\nabla_a n_b,\qquad I_{GH}=\frac{1}{8\pi G}\int\, d^2x \sqrt{h} \text{Tr}K ,\qquad  I_{B}=-\frac{1}{8\pi G}\int\,d^2x \sqrt{h}  $$ }
\begin{eqnarray}
(8\pi G) I_{GH}^{cone}&=& \frac{4\theta_0\cos\epsilon}{\sin^2\epsilon}  \arcsin\left(\tan\epsilon\sinh\tau\right)\large |_{T_1}^{T_2}\nonumber\\  
(8\pi G) I_{B}^{cone}&=&-A^{cone}=- \frac{2\theta_0\cos\epsilon}{\sin^2\epsilon}  \arcsin\left(\tan\epsilon\sinh\tau\right)\large |_{T_1}^{T_2}  
\end{eqnarray}
and the Gibbons-Hawking term and area terms for the constant $\tau$ surfaces ,
 \begin{eqnarray}
(8\pi G) I_{GH}^{\tau}&=& \frac{4\theta_0}{\tan\epsilon} \sinh \tau \sqrt{1-\tan^2\epsilon\sinh^2 \tau}|_{T_1}^{T_2}\nonumber\\  
(8\pi G) I_{B}^{\tau=T}&=&-A^{\tau=T}=- \frac{2\theta_0}{\tan\epsilon} \cosh T\sqrt{1-\tan^2\epsilon\sinh^2 T}
\end{eqnarray}
We immediately see that these last expressions go to zero when $\tau=T$ is taken to reach the curve of the cone at $\sin y=1$ (namely the plane $X=0$), since then $\sinh T=(\tan\epsilon)^{-1}$. In this case the Euclidean action is obtained by 
integrating over the whole (regulated) fundamental region. The resulting volume is $\pi\theta_0 (\tan\epsilon)^{-2} $ as in 
eq.~18 of ref.~\cite{Krasnov}, while the area of the boundary cone is $2\pi\theta_0 \cos\epsilon(\sin\epsilon)^{-2}$ as in 
eq.~19 of that same reference\footnote{In~\cite{Krasnov} the angle $\alpha$ relates to our $\epsilon$ as $\alpha=\pi/2-\epsilon$, and the multiplier $\lambda$ equals $e^{\theta_0}$.}. 

However, there is an important point to consider now: we want to fix the complex/conformal structure at the initial and final surfaces, which implies that in order to have a well-defined variational principle, the Gibbons-Hawking term should be multiplied by 1/2 \cite{Anderson:2010ph,Witten:2018lgb}. Moreover, there is no a priori reason to include the area terms 
$I_B$, so we discard them. The result is then :
\begin{equation}\label{onshellactionceromu}
(-8\pi G)I \simeq -2\theta_0 \arcsin\left(\tan\epsilon\sinh\tau\right)\large |_{T_1}^{T_2} 
\end{equation}
We see that in the $\epsilon\rightarrow 0$ limit we get zero. We also get Krasnov's result \cite{Krasnov} if we let again 
$-\sinh T_1=\sinh T_2=(\tan\epsilon)^{-1}$, even though we do not include area terms. This can be seen as an 
improvement in the way one should understand the computations of \cite{Krasnov} and \cite{Takhtajan:2002cc}: the area 
terms in those works were introduced simply to get finite results, however those terms are metric-dependent, in contrast to 
the more natural conformal-structure dependence we use here, which only needs (one-half of) Gibbons-Hawking terms. 
Finally it is worth noticing that the regulating cone is not special at all, we only need to follow\footnote{Notice that we 
are using a notation different from that of \cite{Takhtajan:2002cc}.} Lemma 5.1 in \cite{Takhtajan:2002cc} which says that
 the regulating surfaces at small fixed $\epsilon$ should behave as $Z\sim \epsilon e^{-\phi(X,Y)}+\mathcal{O}(\epsilon^2)$. 
A short computation then shows that for \textit{any conformal factor}  $\phi$, the trace of the second fundamental form is 
$\Tr(K)=2+\mathcal{O}(\epsilon^2)$.  This is enough to get (\ref{onshellactionceromu}). 
 
\subsection*{The case with $\mu\neq0$}
Let us define  $\rho:=y/a$, so the metric (\ref{Euclideanmetric}) reads, 
\begin{eqnarray}
ds^2&=&d\tau^2 + |\cosh\tau \text{sn}(\rho)^{-1}-J  \sinh\tau \text{sn}(\rho)|^2 d\rho^2\nonumber\\
&+& |\cosh\tau \text{sn}(\rho)^{-1}+J  \sinh\tau \text{sn}(\rho)|^2  \left({dx\over a}\right)^2  \nonumber\\
&+& 2i (\bar{J}-J)\cosh\tau \sinh\tau d\rho  {dx\over a}
\end{eqnarray}

The Euclidean action, taking into account the 1/2 factor in front of the GH terms, is
\begin{equation}
I = \left(\frac{-1}{16 \pi G}\right) \left(-4 \int_{\text{I}=(T_1,T_2)} d\tau \int_\mathcal{F} dx\,dy \sqrt{g} +  \int_{\partial( \text{I}\times\mathcal{F})} dz \sqrt{h} \text{Tr}K \right)
\end{equation}
The  regularized range of integration of $y$ at different times is not evident so we have to assume that for 
small arg$(z)=y$ we have
$\mbox{arg}(z)=\epsilon \cosh\tau$. This assumption is motivated by the discussion of the $\mu=0$ case done before, 
but it is ultimately justified
from the embedding of a surface of constant $\tau$ into $\mathbb{H}^3/\Gamma$ \footnote{The embedding of constant $
\tau$ surfaces in the space $\mathbb{H}^3/\Gamma$ for generic $J$ is quite complicated \cite{Workinprogress}. However, for purely imaginary $J$ the expressions simplify enormously (see the Appendix), and it is indeed the case that the regularizing cone implies arg$(z)=\epsilon \cosh\tau$. Figure \ref{fundamentalregion} was plotted using such embedding.}.  

The volume part is:
\begin{equation}
(16 \pi G) I_{EH} \simeq\frac{-8\theta_0}{a \epsilon} \sinh\tau|_{T_1}^{T_2} + \frac{4\theta_0}{a} \left(E(-|J|^2)-K(-|J|^2)\right) \sinh 2\tau|_{T_1}^{T_2} .
\end{equation}
The one-half Gibbons-Hawking term coming from the fixed time surfaces is
\begin{equation}\label{timeGH}
{16 \pi G\over 2} I_{GH-time}\simeq \frac{4\theta_0}{a \epsilon} \sinh\tau - \frac{4\theta_0}{a} \left(E(-|J|^2)-K(-|J|^2)\right) \sinh 2\tau|_{T_1}^{T_2} .
\end{equation}
We already see that the finite-term contributions cancel when we sum the bulk part plus the GH of constant-time surfaces. In contrast, the divergent part is still there, because we need to consider the rest of the boundary, namely the regulating surface contribution. 

Let us consider again the cone in $\mathbb{H}^3$, which in canonical coordinates is defined by \\ 
$Z/\sqrt{X^2+Y^2+Z^2}=\sin\epsilon$. This means that, for small $\epsilon$, arg$(z)\simeq \epsilon\cosh\tau $. The induced metric, in the $(\tau,x)$ coordinates, can be obtained from the embedding in the Appendix, 
\begin{equation}
ds^ 2|_{cone} \simeq \left[\cosh^2\tau + \mathcal{O}(\epsilon^2)\right] d\tau^2 + \left[\epsilon^{-2}+ \mathcal{O}(\epsilon^0)\right] \theta_0^2 d(x/a)^2 +\mathcal{O}(\epsilon)d\tau dx/a .
\end{equation} 
This implies that the square root of the determinant is $\sqrt{h}\simeq \cosh\tau (\theta_0/a)\epsilon^{-1}+\mathcal{O}(\epsilon)$.  The normal to the cone is $n=-\epsilon^{-1} d\epsilon$, which means Tr$K=\cos\epsilon+\sec\epsilon \simeq 2 + \mathcal{O}(\epsilon^2)$. We can now compute the GH radial term:
\begin{equation}\label{radialGH}
(16 \pi G) (1/2) I_{GH-cone}\simeq \frac{4\theta_0}{a \epsilon} \sinh\tau|_{T_1}^{T_2} + \mathcal{O}(\epsilon) .
\end{equation} 
This term gives the exact same divergent contribution as the GH terms~(\ref{timeGH}) at fixed-time surfaces, which help cancelling the $1/\epsilon$ divergence of the volume part. 

Then, the on-shell action reads
\begin{equation}
I\simeq \mathcal{O}(\epsilon) .
\end{equation}


\subsection{Summing over the instanton moduli space}
\label{modulispacesum}

We have seen that, by equation~(\ref{A}), there are curves in the complex $J$ plane (actually inside the disk $|J|<\coth\tau$) that are defined by the pair $(\theta_0,\theta)$ (here $\theta$ should be thought as independent of $\theta_0$). These level curves were plotted in Figure \ref{contourgrid}.  

\begin{figure}[h!]
\centering
    \includegraphics[width=\linewidth]{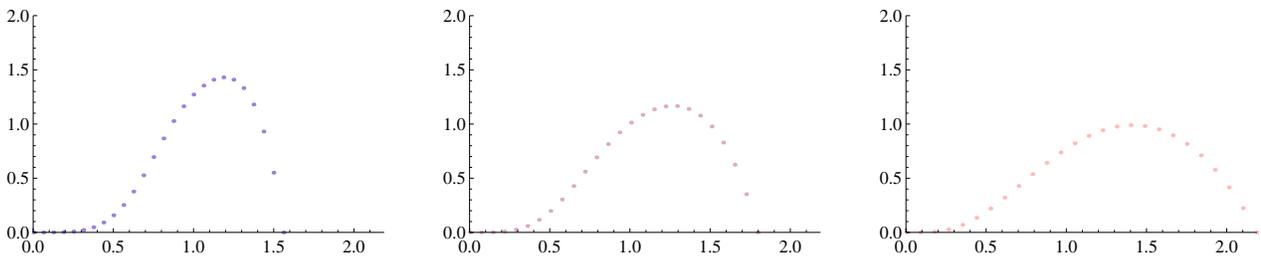} \\
\caption{ Three separate plots of the probability density vs $\theta/\theta_0$ each for different times. From left to right $\tau=.14, 0.34, 4.1$.}
\label{evolutionofprobability}
\end{figure}

Each level curve denotes a set of instantons that interpolate between the given initial and final complex structures. Having 
shown in the previous Section that the action of any such instanton is zero,  we reach a first conclusion: processes
that change the aspect ratio of the cylinder are in fact classically allowed. This agrees with the result of the analysis of
the exact quantum evolution in the ``constrain first'' approach.
The vanishing of the classical instanton action also means that the contribution to the path integral is just given by the 
integration measure. This object should be defined in a sensible way. For the moment, it seems reasonable to just take the Euclidean length $l_\tau(\theta,\theta_0)$ of the level curve as a good measure, although we will only use it for numerical 
evaluation\footnote{It is of course worth exploring the possibility of choosing other measures, but we leave this to future 
work.}. Then 
\begin{equation}
|\langle \theta,\tau | \theta_0,0\rangle | = l_\tau(\theta,\theta_0) 
\end{equation}
Although we are not indicating it explicitly, these lengths are a function of $\theta/\theta_0$. This means that the length (path integral) should be invariant under scaling $\theta \rightarrow b\,\theta $, for any real and positive $b$. This allows us to rewrite the previous equation as 
\begin{equation}
|\langle \theta,\tau |\theta_0,0\rangle | = l_\tau\left(\frac{\theta}{\theta_0}\right) .
\end{equation} 
To interpret this expression as a probability amplitude we should impose that the length is
 normalized with a scale-invariant probability measure, i.e.
\begin{equation}
\int_0^\infty \frac{d\theta}{\theta} |\langle \theta,\tau |\theta_0,0\rangle |^2=1,\qquad \forall\, \theta_0\in\mathbb{R}_+,\quad \tau\in \mathbb{R}
\end{equation}
By a numerical approximation it is possible to plot the normalized length, $|\langle \theta,\tau |\theta_0,0\rangle |^2$, as a function of $\theta/\theta_0$, see Figure \ref{evolutionofprobability}.  

In the coordinate $x:=\log \theta$ the measure of integration
      $d\theta/\theta$ becomes $dx$, the only translation invariant measure
      in $\mathbb{R}$. The Hilbert space is of course 
      $\mathcal{H}=\mathcal{L}^2(\mathbb{R},x)$.


\section{Conclusions}

We studied what is perhaps the simplest example of transition between 2D metrics in pure $AdS3$ quantum gravity, 
to wit: transitions between cylinders with different aspect ratios. In this setting topology does not change and the very 
notion of time evolution is not unique. 

We first defined the evolution in classical gravity, in a set of unconstrained 
coordinates. We exhibited different sets of canonical coordinates together with their respective Hamiltonians. All 
Hamiltonians were obtained via time-dependent canonical transformations starting from a set of canonical coordinates 
in which the Hamiltonian was $H=0$. The fact that time evolution can be trivialized by a choice of canonical coordinates
is of course a general property of classical mechanics. What makes gravity ``special'' is not that, but rather the difficulty
of finding a rationale for a specific nonzero Hamiltonian. Our choice was motivated by the request to mimic a geometric
evolution between cylinder topologies of different aspect ratios. After quantization the time-dependent Sch\"roedinger 
equation could be solved exactly in the momentum representation, where the momentum in this case was identified with
the coordinate canonically conjugate to the aspect ratio.

We changed gear next, by studying the evolution between geometries using the semiclassical Euclidean path integral 
formulation. The main result here is that not all aspect ratios can be reached from evolution starting from a given initial 
ratio, but that for those that can there is no semiclassical barrier, since the classical instanton for the process vanishes. We also proposed
a measure for the wave function obtained from path integral quantization.

Before discussing the relation between the results of the two approaches let us summarize again in some more details
our findings

\begin{enumerate}

\item The classical phase space of a cylinder topology depends on two real numbers, one measuring the aspect of the cylinder, the other being a kind of momentum. There are many possible choices of such pairs, and we have identified two relevant ones together with their corresponding Hamiltonian. In particular we paid attention to the case where the generalized coordinate is the aspect and we showed how it evolves in time. 

\item Fixing the initial complex structure by some $\lambda_0 \in (1,\infty)$, the saddle solutions are able to reach only a final $\lambda_\tau$ in the interval $(1,\lambda_0^M)$, where the real number $M$ depends on the equidistant time $\tau$ elapsed between initial and final complex structures. For $\tau\rightarrow\infty$ the interval is $(1,\lambda_0^{2.18844})$. All other complex structures have probability zero to be reached (by means of these saddles).

\item There are infinite interpolating saddles.

\item All the saddles have on-shell action equal to zero. Then the only difference in their contributions to the path-integral giving $\langle \theta_\tau , \tau | \theta_0,0\rangle $ comes from the integration measure $\int \text{d(saddles)}$, which sweeps the different saddles (more precisely a curve in the space of holomorphic quadratic differentials, see Section \ref{modulispacesum}). After choosing such a measure we get Figure \ref{evolutionofprobability}.

\item  Since the probability $|\langle \theta,\tau | \theta_0,0\rangle|^2$ is a function of $\theta/\theta_0$, there is a natural Hilbert space given by $\mathcal{H}=\mathcal{L}^2\left(\mathbb{R}_+,\frac{d\theta}{\theta}\right)$. In this representation $|\theta\rangle=\theta\delta_\theta$ (i.e., $\theta$ times the delta centered at $\theta$). 

\item The range of possible final complex structures of point 2 above is more stringent that the purely geometric result of Proposition 3 in \cite{Gardiner}\footnote{See also example 3 and theorem 3 of \cite{Ahlfors}}: $K^{-1}\log\lambda_0\leq \log\lambda_\tau\leq K \log\lambda_0 $, where $K=(1+|J|\tanh\tau )/(1-|J|\tanh\tau)$ is the bound of the corresponding $K$-quasiconformal map (\ref{quasimap}) deforming the Riemann surface with $\theta_0$ to the Riemann  surface with $\theta_\tau$ (and Beltrami differential given by $\mu=J \tanh\tau \text{sn}^2(y/a)$ ).     

\item The integration measure over the moduli of instantons interpolating between the two complex structures is chosen as follows: i) the value $\theta/\theta_0$ fixes a curve in the space of quadratic differentials $q=\bar{J}/(a^2) dw^2$. ii) this curve in the complex disk  $|J|<\coth\tau$  has \textit{Euclidean} length $l$. iii) Define then $\int d(\text{saddles})=l$. This is developed in Section \ref{modulispacesum}. It would be interesting to understand if the measure can be obtained from a 
``natural" measure on the space of holomorphic quadratic differentials. 
 
\end{enumerate}

Let us conclude by spending a few words on time evolution in the two approaches described
in this paper. As we mentioned earlier, the path integral evolution should agree with the exact quantum evolution to lowest
order in the semiclassical expansion; it is therefore interesting to see how the two are related. 
One of the results of Section 3 is 
that the transition between different aspect ratios $\lambda$ has no semiclassical barrier --since the corresponding 
instanton action vanishes. The action vanishes for the specific choice of boundary terms that keeps the initial and 
final metrics fixed up to a conformal factor. This choice of data is appropriate for computing transitions between eigenstates 
of the canonical coordinate $Q$.
 The solution of the Schr\"odinger equation shown in~(\ref{mm9}) computes instead transitions between eigenstates of the 
 conjugate momentum $P$ and the time evolution~(\ref{mm9}) carries a $P$-eigenstate into an eigenstate of the time-evolved momentum, without any
 additional phase shift
 \beq
 |P\rangle \rightarrow | \hat{P}(P,t)\rangle.
 \eeq{mmm2}
So, we arrive at the curious result that ~(\ref{mm9}) makes eigenstates of $P$ evolve without phase shift 
while the semiclassical evolution studied in Section 3 seems to do the same for $Q$ eigenstates. This puzzle is resolved by 
noticing that the same choice of boundary term in the Euclidean action that fixes $Q$ eigenstates, namely $(1/2)I_{GH}$, 
makes the action stationary also under variations that hold the traceless part of the extrinsic curvature fixed. 
This is seen by writing the action $I_{EH}+(1/2)I_{GH}$ in ADM variables and using the same notations as in eq.~(\ref{sing}) 
\beq
 I_{EH}+{1\over 2}I_{GH}={1\over 16\pi G} \int dt \int d^2x \sqrt{h} N[K_{ab}K^{ab} -K^2] +{1\over 16\pi G} \int d^2x \sqrt{h} K
 +... .
 \eeq{var1}
The variation of the terms denoted by $....$ does not produce any boundary terms while the rest gives the terms
\beq
\delta I_{EH}+{1\over 2}\delta I_{GH}= {1\over 16\pi G} \int d^2x \sqrt{h} \left(\delta h_{ab} K^{ab} -\delta h_{ab} h^{ab} K + \delta K + 
{K\over 2}\delta h_{ab}h^{ab} \right) .
\eeq{var2}
This equation shows that $I_{EH} +I_{GH}$ is stationary under the variation $\delta h_{ab}=\omega h_{ab}$, 
$\delta K=0$, as in~\cite{Witten:2018lgb}, but {\em also} under an arbitrary variation of the boundary 
metric, if the traceless part of $K_{ab}$ is kept  fixed and $K$ changes as 
\beq
\delta K = -\delta h_{ab} K^{ab}  + {K\over 2}\delta h_{ab}h^{ab}.
\eeq{var3}
So, at the lowest order in the semiclassical approximation, the action we used can also describe transitions between $P$
eigenstates. It is therefore comforting to see that it gives an evolution compatible with eq.~(\ref{mmm2}).

\subsection*{Acknowledgements}
We would like to thank G. Giribet, M. Leston, M. Mereb and S. Wolpert for helpful comments. AG is supported in part by PICT2016 0094 grant.  MP is supported in part by NSF grant PHY-1915219.
\setcounter{equation}{0}
\renewcommand{\theequation}{A.\arabic{equation}} 
\section*{Appendix}
\begin{appendix}

Here we describe an embedding of surfaces of constant $\tau$ into the hyperbolic space $\mathbb{H}^3/\Gamma$. 
Actually, we will only consider the case where $J$ is purely imaginary: $J=i k$, with $k\in \mathbb{R}$.  The equations to 
solve in the generic case are much more complicated and will be discussed elsewhere \cite{Workinprogress}.

Let us call $Z$ and $W$, respectively,  the real positive and complex coordinates of 
$\mathbb{H}^3/\Gamma$ respectively, such that $ds^2=(dZ^2+|dW|^2)/Z^2$. The embedding is  given by
\begin{eqnarray}
\label{Equidembedding}
W&=&e^ {(1+i k) {x\over a}} \left(\frac{\text{cn}\rho \cosh\tau - i \text{sn}\rho \text{dn}\rho\sinh\tau} {\text{dn}\rho 
\cosh\tau - k \text{sn}\rho \text{cn}\rho\sinh\tau} \right) ,\nonumber\\  
Z&=&\frac{\sqrt{1+k^2}|\text{sn}\rho| e^{x\over a}}{\text{dn}\rho \cosh\tau - k \text{sn}\rho \text{cn}\rho\sinh\tau} .
\end{eqnarray}
The Jacobi elliptic functions have parameter $-k^2$ and we set $\rho=y/a$. Note that the regularizing cone given by 
$Z/|W|\simeq \epsilon$ implies $\text{sn}\rho\simeq \rho\simeq \epsilon \cosh\tau $ after absorbing irrelevant factors into 
$\epsilon$.

\end{appendix}


\begin{thebibliography}{99}
\bibitem{deser}
S.~Deser and R.~Jackiw,
  ``Three-Dimensional Cosmological Gravity: Dynamics of Constant Curvature,''
  Annals Phys.\  {\bf 153}, 405 (1984);
S.~Deser, R.~Jackiw and G.~'t Hooft,
  ``Three-Dimensional Einstein Gravity: Dynamics of Flat Space,''
  Annals Phys.\  {\bf 152}, 220 (1984);

\bibitem{Banados:1992wn} 
  M.~Banados, C.~Teitelboim and J.~Zanelli,
  ``The Black hole in three-dimensional space-time,''
  Phys.\ Rev.\ Lett.\  {\bf 69}, 1849 (1992)
  
\bibitem{Banados:1992gq} 
  M.~Banados, M.~Henneaux, C.~Teitelboim and J.~Zanelli,
  ``Geometry of the (2+1) black hole,''
  Phys.\ Rev.\ D {\bf 48}, 1506 (1993)
  Erratum: [Phys.\ Rev.\ D {\bf 88}, 069902 (2013)]
  


\bibitem{Brill:1995jv} 
  D.~R.~Brill,
  ``Multi - black hole geometries in (2+1)-dimensional gravity,''
  Phys.\ Rev.\ D {\bf 53}, 4133 (1996)
  
\bibitem{Aminneborg:1997pz} 
  S.~Aminneborg, I.~Bengtsson, D.~Brill, S.~Holst and P.~Peldan,
  ``Black holes and wormholes in (2+1)-dimensions,''
  Class.\ Quant.\ Grav.\  {\bf 15}, 627 (1998)
  
\bibitem{Brill:1998pr} 
  D.~Brill,
  ``Black holes and wormholes in (2+1)-dimensions,''
  Lect.\ Notes Phys.\  {\bf 537}, 143 (2000)
  [gr-qc/9904083].

\bibitem{Martinec:1984fs} 
  E.~J.~Martinec,
  ``Soluble Systems in Quantum Gravity,''
  Phys.\ Rev.\ D {\bf 30}, 1198 (1984).
  
  
\bibitem{Achucarro:1987vz} 
  A.~Achucarro and P.~K.~Townsend,
  ``A Chern-Simons Action for Three-Dimensional anti-De Sitter Supergravity Theories,''
  Phys.\ Lett.\ B {\bf 180}, 89 (1986).
  
\bibitem{Witten:1988hc} 
  E.~Witten,
  ``(2+1)-Dimensional Gravity as an Exactly Soluble System,''
  Nucl.\ Phys.\ B {\bf 311}, 46 (1988).
  
\bibitem{Witten:1989sx} 
  E.~Witten,
  ``Topology Changing Amplitudes in (2+1)-Dimensional Gravity,''
  Nucl.\ Phys.\ B {\bf 323}, 113 (1989).
  
\bibitem{Martin:1989mb} 
  S.~P.~Martin,
  ``Observables in (2+1)-dimensional Gravity,''
  Nucl.\ Phys.\ B {\bf 327}, 178 (1989).
  
\bibitem{Carlip:1989nz} 
  S.~Carlip,
  ``Exact Quantum Scattering in (2+1)-Dimensional Gravity,''
  Nucl.\ Phys.\ B {\bf 324}, 106 (1989).    
 
 
\bibitem{Moncrief:1989dx} 
  V.~Moncrief,
  ``Reduction of the Einstein equations in (2+1)-dimensions to a Hamiltonian system over Teichmuller space,''
  J.\ Math.\ Phys.\  {\bf 30}, 2907 (1989).

\bibitem{Hosoya:1989yj} 
  A.~Hosoya and K.~i.~Nakao,
  ``(2+1)-dimensional Pure Gravity for an Arbitrary Closed Initial Surface,''
  Class.\ Quant.\ Grav.\  {\bf 7}, 163 (1990).

\bibitem{Carlip}
  S.~Carlip,
  ``Observables, Gauge Invariance, and Time in (2+1)-dimensional
  Quantum Gravity,''
  Phys.\ Rev.\ D {\bf 42}, 2647 (1990).

\bibitem{kp15} 
  J.~Kim and M.~Porrati,
  ``On a Canonical Quantization of 3D Anti de Sitter Pure Gravity,''
  JHEP {\bf 1510}, 096 (2015)
  doi:10.1007/JHEP10(2015)096
  [arXiv:1508.03638 [hep-th]].

\bibitem{Scarinci:2011np} 
  C.~Scarinci and K.~Krasnov,
  ``The universal phase space of $AdS_3$ gravity,''
  Commun.\ Math.\ Phys.\  {\bf 322}, 167 (2013)
  doi:10.1007/s00220-012-1655-0
  [arXiv:1111.6507 [hep-th]].

\bibitem{Krasnov:2005dm} 
  K.~Krasnov and J.~M.~Schlenker,
  ``Minimal surfaces and particles in 3-manifolds,''
  Geom.\ Dedicata {\bf 126}, 187 (2007)


\bibitem{Nag} S. Nag, ``The complex analytic theory of Teichm\"uller spaces'', John
Wiley, New York, Chichester, Brisbane, Toronto, Singapore (Canadian Mathematical Society Series of Monographs and Advanced Texts),
1988,  ISBN 0-471-62773-9 .



\bibitem{Gardiner}  F. P. Gardiner, \textit{Quasiconformal Teichm\"uller Theory}, American Mathematical Society (2000).

\bibitem{arnold} V.I. Arnold, 
{\em Geometrical Methods in the Theory of Ordinary Differential Equations}, Springer (2012).

\bibitem{haha} 
  J.~B.~Hartle and S.~W.~Hawking,
  ``Wave Function of the Universe,''
  Phys.\ Rev.\ D {\bf 28}, 2960 (1983)
  [Adv.\ Ser.\ Astrophys.\ Cosmol.\  {\bf 3}, 174 (1987)].
  
\bibitem{Ahlfors} L. V. Ahlfors, ``Lectures on quasiconformal mappings'' , University Lecture Series, vol. 38 (2006).
 
 
\bibitem{Krasnov} 
  K.~Krasnov,
  ``Holography and Riemann surfaces,''
  Adv.\ Theor.\ Math.\ Phys.\  {\bf 4}, 929 (2000)


\bibitem{Takhtajan:2002cc} 
  L.~A.~Takhtajan and L.~P.~Teo,
  ``Liouville action and Weil-Petersson metric on deformation spaces, global Kleinian reciprocity and holography,''
  Commun.\ Math.\ Phys.\  {\bf 239}, 183 (2003)

\bibitem{Anderson:2010ph} 
  M.~T.~Anderson,
  ``On quasi-local Hamiltonians in General Relativity,''
  Phys.\ Rev.\ D {\bf 82}, 084044 (2010)

\bibitem{Witten:2018lgb} 
  E.~Witten,
  ``A Note On Boundary Conditions In Euclidean Gravity,''
  arXiv:1805.11559 [hep-th].

\bibitem{Workinprogress} A. Garbarz, J. Kim and M. Porrati, Work in progress.

\end{thebibliography}
 \end{document}